\documentclass{aa}  

\usepackage{graphicx}
\usepackage{natbib}
\bibpunct{(}{)}{;}{a}{}{,} 

\setcitestyle{authoryear,open={(},close={)}} 

\def \feix {Fe\,{\sc ix}}
\def \fexv {Fe\,{\sc xv}}
\def \fexix {Fe\,{\sc xix}}
\def \fexxi {Fe\,{\sc xxi}}
\def \fexxiii {Fe\,{\sc xxiii}}

\begin{document}

   \title{Coronal energy release by MHD avalanches}

   \subtitle{II. EUV line emission from a multi-threaded coronal loop}

   \author{G. Cozzo\inst{1}
          \and
          J. Reid \inst{2}
          \and
          P. Pagano \inst{1,3}
          \and
          F. Reale \inst{1,3}
          \and
          P. Testa \inst{4}
          \and
          A.~W. Hood \inst{2}
          \and
          C. Argiroffi \inst{1,3}
          \and
          A. Petralia \inst{3}
          \and
          E. Alaimo \inst{1}
          \and
          F. D'Anca \inst{3}
          \and
          L. Sciortino \inst{3}
          \and
          M. Todaro \inst{3}
          \and 
          U. Lo Cicero \inst{3}
          \and
          M. Barbera \inst{1,3}
          \and
          B. de Pontieu \inst{5,6,7}
          \and
          J. Martinez-Sykora \inst{5,8,6,7}
          }

   \institute{
            Dipartimento di Fisica \& Chimica, Università di Palermo, Piazza del Parlamento 1, I-90134 Palermo, Italy
            \email{gabriele.cozzo@unipa.it}
            \and School of Mathematics and Statistics, University of St Andrews, St Andrews, Fife, KY16 9SS, UK
            \and INAF-Osservatorio Astronomico di Palermo, Piazza del Parlamento 1, I-90134 Palermo, Italy
            \and Harvard–Smithsonian Center for Astrophysics, 60 Garden St., Cambridge, MA 02193, USA 
            \and Lockheed Martin Solar \& Astrophysics Laboratory, 3251 Hanover St, Palo Alto, CA 94304, USA
            \and Rosseland Centre for Solar Physics, University of Oslo, P.O. Box 1029 Blindern, N-0315 Oslo, Norway
            \and Institute of Theoretical Astrophysics, University of Oslo, P.O. Box 1029 Blindern, N-0315 Oslo, Norway
            \and Bay Area Environmental Research Institute, NASA Research Park, Moffett Field, CA 94035, USA
            }
            
   \date{Received \dots ; accepted \dots}

 
  \abstract
  {Magnetohydrodynamic (MHD) instabilities, such as the kink instability, can trigger the chaotic fragmentation of a twisted magnetic flux tube into small-scale current sheets that dissipate as aperiodic impulsive heating events. 
  In turn, the instability could propagate as an avalanche to nearby flux tubes and lead to a nanoflare storm. Our previous work was devoted to related 3D MHD numerical modeling, which included a stratified atmosphere from the solar chromosphere to the corona, tapering magnetic field, and solar gravity for curved loops with the thermal structure modelled by plasma thermal conduction, along with optically thin radiation and anomalous resistivity for 50 Mm flux tubes.}
  {Using 3D MHD modeling, this work addresses predictions for the extreme-ultraviolet (EUV) imaging spectroscopy of such structure and evolution of a loop, with an average temperature of 2-2.5 MK in the solar corona. We set a particular focus on the forthcoming MUSE mission, as derived from the 3D MHD modeling.}
  {From the output of the numerical simulations, we synthesized the intensities, Doppler shifts, and non-thermal line broadening in 3 EUV spectral lines in the MUSE passbands: \feix\ 171\AA, \fexv\ 284\AA, and \fexix\ 108\AA, emitted by  $\sim 1\,\mathrm{MK}$, $\sim 2\,\mathrm{MK}$, and $\sim 10\,\mathrm{MK}$ plasma, respectively. These data were detectable by MUSE, according to the MUSE expected pixel size, temporal resolution, and temperature response functions.  We provide maps showing different view angles (front and top) and realistic spectra. Finally, we discuss the relevant evolutionary processes from the perspective of possible observations.} 
  {We find that the MUSE observations might be able to detect the fine structure determined by tube fragmentation. In particular, the \feix\ line is mostly emitted at the loop footpoints, where we might be able to track the motions that drive the magnetic stressing and detect the upward motion of evaporating plasma from the chromosphere. In \fexv, we might see the bulk of the loop with increasing intensity, with alternating filamentary Doppler and non-thermal components in the front view, along with more defined spots in the topward view. The \fexix\ line is very faint within the chosen simulation parameters; thus, any transient brightening around the loop apex may possibly be emphasized by the folding of sheet-like structures, mainly at the boundary of unstable tubes.}
  {In conclusion, we show  that  coronal loop observations with MUSE can pinpoint some crucial features of MHD-modeled ignition processes, such as the related dynamics, helping to identify the heating processes.}

\maketitle
\section{Introduction}

\begin{figure*}[h!]
  \includegraphics[width=1.\hsize]{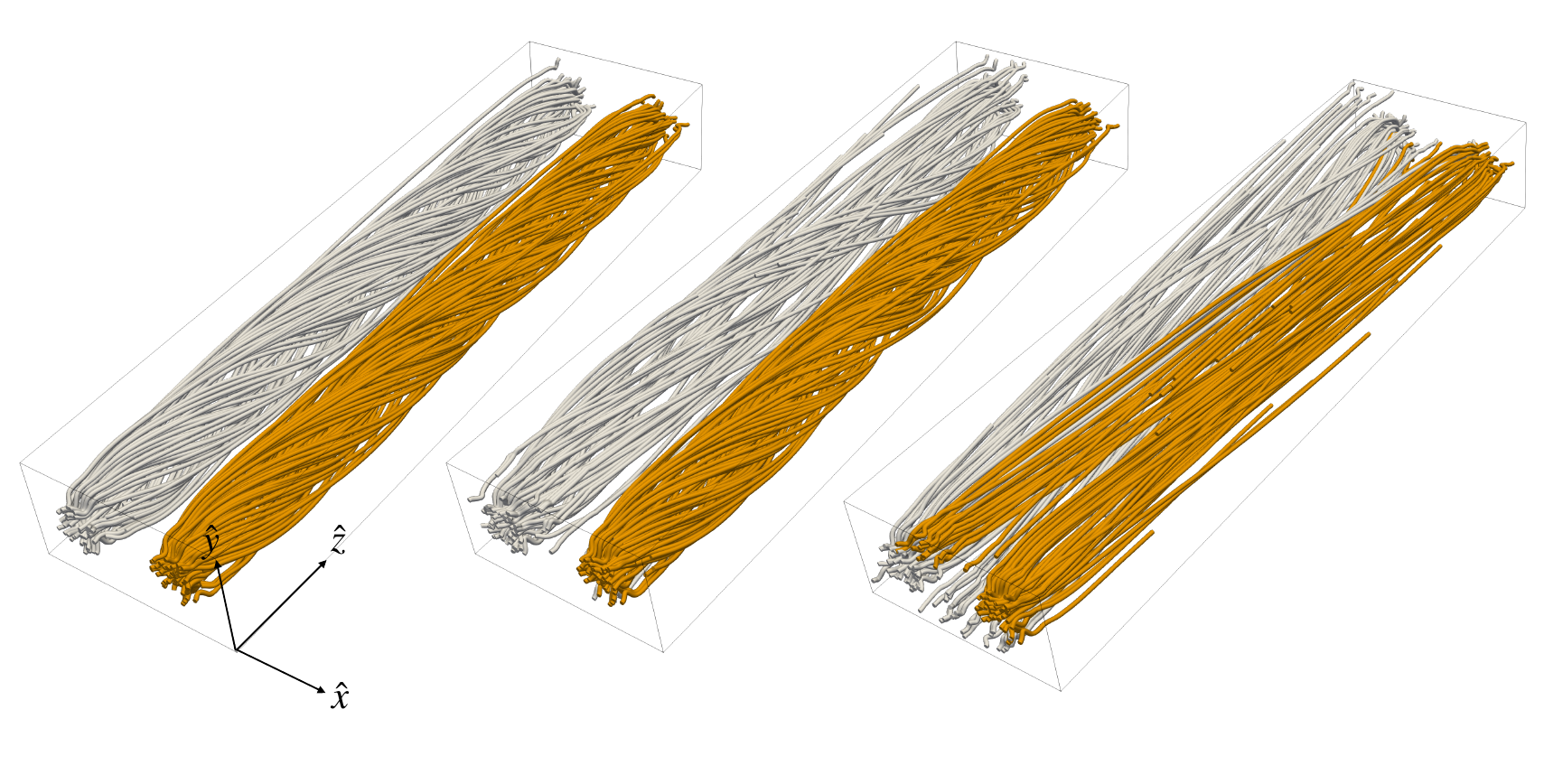}
  \caption{3D rendering of the magnetic field lines inside the box at times (from left to right): $t = 0\,\mathrm{s}$ (initial condition), $180\,\mathrm{s}$, (first loop disruption), and $500\,\mathrm{s}$. The change in the field line connectivity during the evolution of the MHD cascade is emphasized by the colors. }
  \label{Fig:avalance}
\end{figure*}

\begin{figure*}[h!]
   \sidecaption
   \includegraphics[width=12cm]{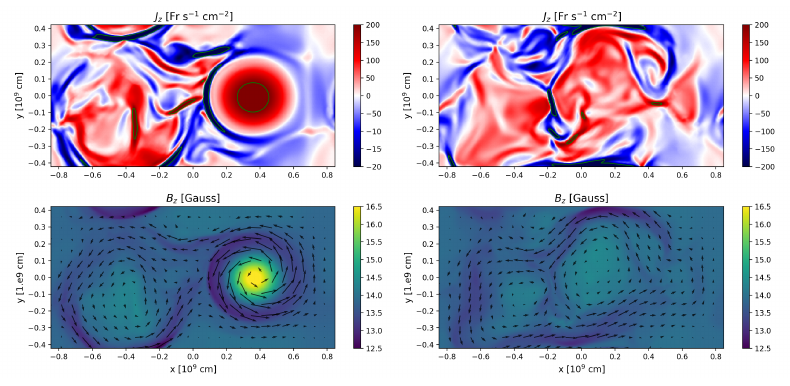}
  \caption{MHD avalanche 2D snapshots. \textbf{Upper panels.} Horizontal cuts of the current density across the mid plane at time $t = 180\,\mathrm{s}$ (left) and $t = 285\,\mathrm{s}$ (right). The green contours encloses the regions of the domain heated by ohmic dissipation. \textbf{Lower panels.} Horizontal cuts of the magnetic field across the mid-plane at the same snapshot-times as before. The vector field is included.}
  \label{Fig:cut_1}
\end{figure*}

The solar corona consists of plasma confined by and interacting with the coronal magnetic field \citep{vaiana1973identification, testareale2023xray}. Frequently, it appears as arch-shaped, dense, and bright filamentary structures, confined by the magnetic field, and known as coronal loops \citep[e.g.,][]{reale2014coronal}. These loops are at temperatures above one million Kelvin, however, the explanation behind this still remains a long-standing problem \citep{grotrian1939sonne, edlen1943deutung, peter2014discovery}.

Presently, there is a general acknowledgment of the magnetic field's predominant role as the primary source of heating energy (\citealt{Alfven1947,parker1988nanoflares}, and see also reviews by e.g., \citealt{klimchuk2015key,testareale2023xray} and references therein).
In particular, to address the coronal heating problem, two main classes of mechanisms have been envisaged: one involves the dissipation of stored magnetic stresses, referred to as DC heating and the other involves the damping of magnetohydrodynamic (MHD) waves, known as AC heating \citep{parnell2012contemporary, zirker1993coronal}. 

Overall, DC heating involves magnetic energy storage and impulsive, widespread release. In fact, as the magnetic field evolves over time, energy builds up in the solar corona, which can later be converted into thermal energy. In particular, turbulent photospheric motions cause coronal magnetic field lines to twist and tangle with each other, leading to the inevitable growth of magnetic stresses \citep{parker1988nanoflares, klimchuk2015key}. 
As a consequence of the ongoing stirring of plasma in the photosphere, caused by magnetoconvection, the magnetic field lines (forming coronal loops) are expected to exhibit intricate braiding patterns at extremely fine resolutions, smaller than an arcsecond \citep{klimchuk2009coronal, cirtain2013energy}. Parker envisaged that this ongoing process would ultimately give rise to widespread formation of tangential discontinuities in the magnetic field and small-scale current sheets within the solar corona. These discontinuities would serve as sites for magnetic reconnection events, leading to the release of small amounts of energy, on the order of $10^{24}\,\mathrm{erg}$ in what are called \lq nanoflares\rq\ \citep{parker1988nanoflares}.

Although photospheric plasma induces slow and local dragging of magnetic field lines at loop "footpoints," according to Parker's theory, magnetic energy release in the large scale coronal environment is expected to occur through impulsive and widespread heating events.
One potential heating mechanism that can produce rapid surges of energy release from the slow, continual driving of boundary motions is the avalanche model \citep{2016mhhoodd}.
In this model, a localized MHD instability, within a single strand of a coronal loop, results in widespread disruption as neighboring strands become affected by the propagating disturbance.
This mechanism offers a promising explanation for the complex interplay between slow and fast processes in coronal energy release.

In particular, as turbulent motion induces twisting of the loop's magnetic field lines, such flux tubes can become susceptible to the kink instability \citep{hood1979kink, hood2009coronal}, resulting in the release of magnetic energy through sudden, widespread heating events. The initial helical current sheet progressively fragments in a turbulent way into smaller scale sheets. The turbulent dissipation of the magnetic structure into small-scale current sheets occurs as a sequence of aperiodic, impulsive heating events, similar to nanoflare storms.  

This is the case of the numerical experiments described in \cite{tam2015coronal}, \cite{2016mhhoodd}, and \cite{reid2018coronal, reid2020coronal}, where strongly twisted multi-threaded coronal loops undergo kink instabilities and dissipate energy through a turbulent cascade of current sheets. 
In these cases, purely coronal loops were taken into account, without considering the interaction with the underlying chromospheric layer.
In particular, when multiple magnetic strands coexist within the same coronal loop, a single unstable magnetic strand could disrupt the  stable strands nearby and trigger a global dissipation of magnetic energy, through a domino-effect called an "MHD avalanche"\ \citep{2016mhhoodd}.
Recently, \cite{cozzo2023coronal} addressed the problem of MHD avalanches by means of non-ideal MHD modeling and fully 3D numerical simulations.
They considered a stratified solar atmosphere (including the chromospheric layer, transition region, and lower corona), where multiple interacting coronal loop strands are progressively twisted and eventually become unstable.
They confirm that avalanches are a viable mechanism for the storage and release of magnetic energy in coronal loops, as a result of photospheric motions in a more realistic solar atmosphere.

\begin{table}[h!]
\caption{MUSE spectrometer emission lines.}           
\label{table:1}     
\centering                 
\begin{tabular}{c c c}     
\hline       
\\
Line & Wavelength [$\AA$]  & $\log_{10}$ (T [K])  \\\\
\hline                   
\\
   \feix\ & 171 & 5.9   \\
   \fexv\ & 284 & 6.4 \\
   \fexix\ / \fexxi\ & 108 & 7.0 / 7.1 \\
\\
\hline                                   
\end{tabular}
\end{table}

The Multi-slit Solar Explorer \citep[MUSE,][]{ de2020multi, de2022probing, cheung2022probing} is an upcoming NASA MIDEX mission, featuring a multi-slit extreme-ultraviolet (EUV)  spectrometer and an EUV context imager,   planned for launch in 2027. MUSE is designed to offer high spatial and temporal resolution for spectral and imaging observations of the solar corona. One of the MUSE science goals is to advance the understanding of the heating mechanisms in the corona of both the quiet Sun and active regions, as well as of the physical processes governing dynamic phenomena like flares and eruptions. MUSE will provide fine spatio-temporal coverage of coronal dynamics, as well as wide field of view (FoV) observations, offering valuable insights into the physics of the solar atmosphere. In particular, it will obtain  high resolution spectra ($\approx 0.38"$), with wide angular coverage ($\approx 156" \times 170"$; resembling the typical size of an active region) and $12\,\mathrm{s}$ cadence. 
With its 35-slit spectrometer, MUSE will provide spectral observations, with unprecedented combination of cadence and spatial coverage, in different EUV passbands  dominated by strong lines formed over a wide temperature range, enabling the testing of state-of-the-art models related to coronal heating, solar flares, and coronal mass ejections.

The main EUV lines in the MUSE passbands (171\AA\ \feix, 284\AA\ \fexv, and 108\AA\ \fexix\ and \fexxi), are listed in Table \ref{table:1} (for details, see  \citealt{de2020multi,de2022probing}). Here, we focus on the new spectroscopic diagnostics provided by MUSE, but of course the findings can be extended to observations with other imaging and spectroscopic solar instruments observing at similar wavelengths, for instance, SDO/AIA \citep{pesnell2012solar, lemen2012atmospheric}, Hinode/EIS \citep{culhane2007, tsuneta2008solar}, and the forthcoming Solar-C/EUVST \citep{shimizu2019solar}.

To make meaningful predictions that can then be compared with solar observations, two critical developments are essential. Firstly, the modeling approach must encompass crucial physical components, including the thermodynamic response of atmosphere, to derive realistic observational outcomes.
Secondly, observational techniques must achieve sufficient temporal and spatial resolution within the pertinent spectral bands.
The synergistic comparison between coronal observations and synthetic plasma diagnostics from numerical simulations can, on the one hand, significantly improve our interpretative power on real observational data. On the other hand, it can also help to refine the solar coronal modeling process.

This study focuses on extending the plasma diagnostics of the MHD avalanche model introduced by \cite{cozzo2023coronal} to observations with the forthcoming MUSE spectrograph.
Although  our analysis addresses plasma diagnostics from the MUSE spectrometer, the present work has a more general application to the plasma emission at representative coronal temperatures. 

We extracted plasma diagnostics from a full 3D MHD simulation of a flaring coronal loop undergoing an MHD avalanche, as described in \cite{cozzo2023coronal}. In particular, we modeled the response of the MUSE spectrometer by using our simulated plasma output to synthesize line emission, intensity maps, Doppler shifts and non-thermal line widths, across the three spectral channels available with the instrument.

In Section \ref{sec:model}, the MHD avalanche model from which observables are synthesized is briefly presented.
In Section \ref{sec:diagnostics}, we describe the method employed to extract these synthetic observables. In Section \ref{sec:results}, we present our results. In Section \ref{sec:conclusions}, we discuss the results and give our conclusions.

\section{Model}
\label{sec:model}
As described in \citet{cozzo2023coronal}, we modeled a magnetized solar atmosphere in a 3D cartesian box of size, $-x_M < x < x_M$, $-y_M < y < y_M$, and $-z_M < z < z_M$, where $x_M = 2 y_M = 8.5 \times 10^8\,\mathrm{cm}$, and $z_M = 3.1 \times 10^9\,\mathrm{cm}$ with PLUTO Code \citep{mignone2007pluto}.

The background atmosphere consists of a chromospheric and a coronal column separated by a thin transition region. 
Specifically, in the corona, two flux tubes interact, each with a length $50\,\mathrm{Mm}$ and initial temperature of approximately $10^6\,\mathrm{K}$, while the two chromospheric layers are $\sim 6\,\mathrm{Mm}$ wide each and $10^4\,\mathrm{K}$ hot. The transition region, $\approx 1\,\mathrm{Mm}$ wide, is artificially broadened using Linker--Lionello--Mikić method \citep{linker2001magnetohydrodynamic, lionello2009multispectral, mikic2013importance}. 

The loop's magnetic field is line-tied to the photospheric boundaries, at the opposite sides of the domain. The coronal loop is rooted to the chromosphere by its footpoints and it expands approaching the corona.
Radiative losses and thermal conduction drain internal energy from the corona while a uniform and static heating prevents the atmosphere from cooling down and maintains the corona in steady conditions. 

Two rotational motions at the footpoints twist each of the flux tubes. The two rotating regions have the same radius ($R \approx 1\,\mathrm{Mm}$), but one has an angular velocity that is higher than the other by $10 \%$ ($\approx 10^{-3} \,\mathrm{rad\,s}^{-1}$ vs. $\approx 0.9 \times 10^{-3} \,\mathrm{rad\,s}^{-1}$), so that the faster strand becomes unstable first, triggering the avalanche process.
As a reference time, deemed $t = 0\,\mathrm{s}$, we chose the time when the faster flux tube is nearly kink-unstable. 
Then, this kink-unstable tube rapidly disrupts the other one ($t = 285\,\mathrm{s}$).
A 3D rendering of the two flux tubes and of their interaction as a result of the instability is shown in Fig.\ref{Fig:avalance}. Here, we focus on the evolution after the first kink instability.
 
The instability makes the initially monolithic flux tubes fragment into a turbulent structure of thinner strands.
The initially helical current sheet fragments into smaller and smaller currents sheets, which dissipate by magnetic reconnection and are responsible for a sequence of aperiodic impulsive heating events.

The left panels of Fig.\ \ref{Fig:cut_1} shows a cross-section at the mid-plane of the computational box at time $t=180 \,\mathrm{s}$, after the beginning of the instability, when the flux tube on the left ($x < 0$) has already become unstable and fragmented, and is about to trigger the instability of the flux tube on the right ($x > 0$).
The current density map (top left panel) clearly shows on the left the thin, intense current sheets, around which magnetic reconnection occurs and magnetic energy dissipates.
On the right of the map, we see the other flux tube, still not involved in the instability.
In the lower-left panel, we show the magnetic field intensity: where the left-hand flux tube was, the field is more blurred, dispersed, and irregularly distributed as a consequence of the instability.
The flux tube on the right is still compact and coherent.
The twisting is emphasized by the vector field of the component parallel to the plane.
The right panels of Fig. \ref{Fig:cut_1} show the same quantities as do those on the left, but at a later time ($t = 285\,\mathrm{s}$), when the flux tube on the right has also been disrupted. 

\begin{figure}[h!]
   \centering
  \includegraphics[width=\hsize]{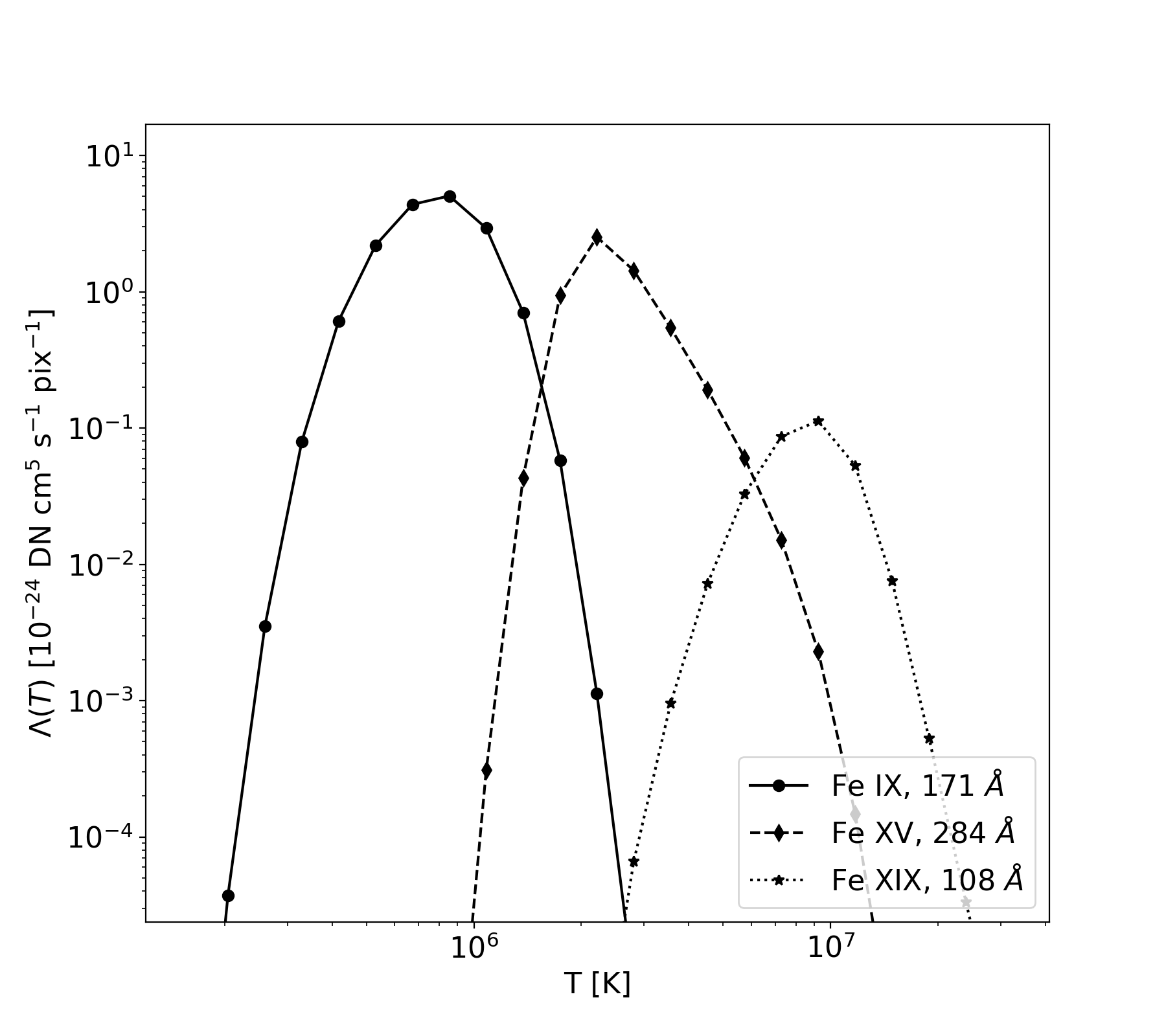}
  \caption{Response function $\Lambda_i (T)$ for the three MUSE emission lines (\feix, \fexv, and \fexix) as a function of temperature.}
  \label{Fig:MandA}
\end{figure}

\section{Methods: Forward modeling}
\label{sec:diagnostics}

\begin{figure*}[h!]
  \includegraphics[width=0.29\hsize]{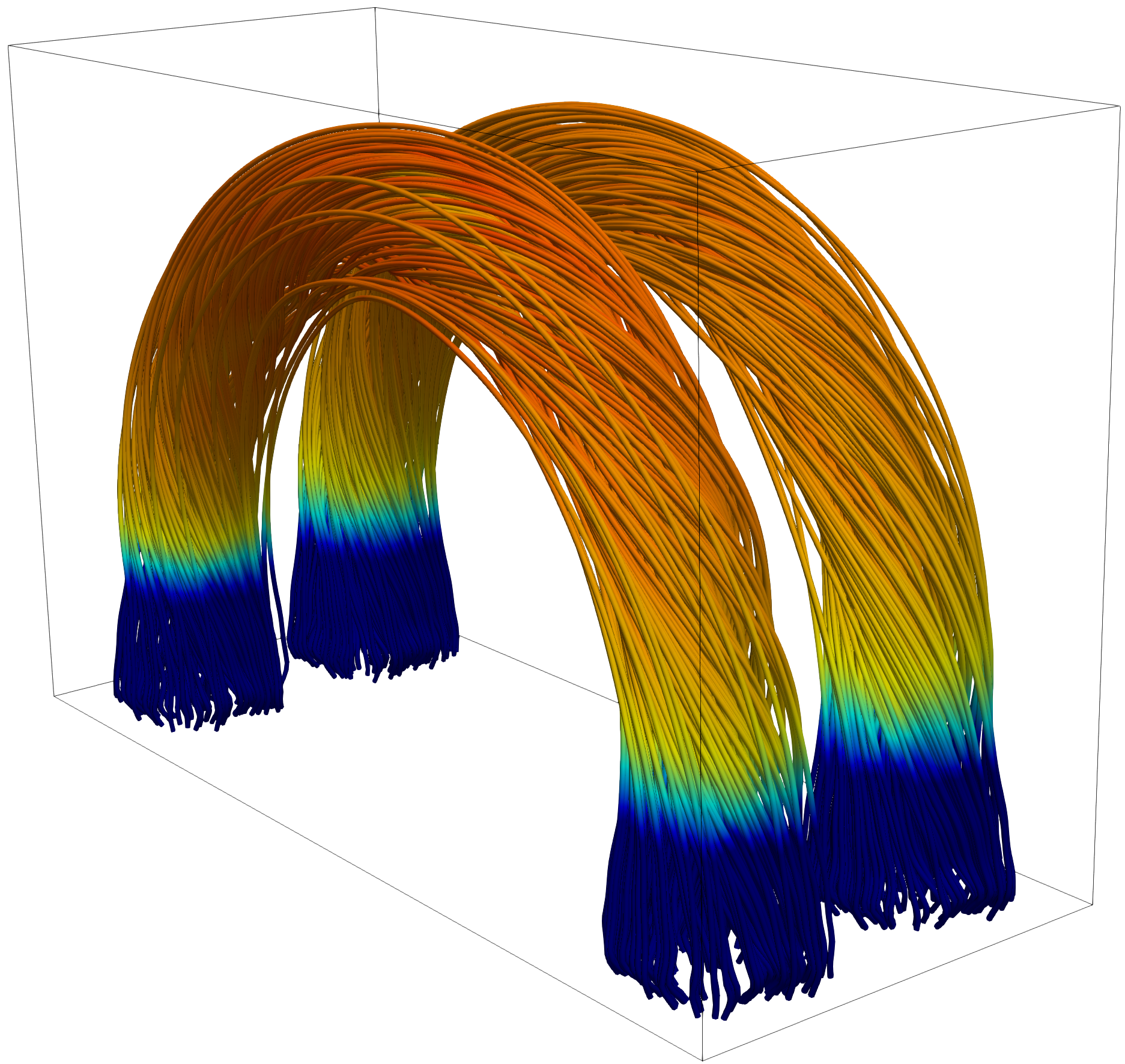}
  \includegraphics[width=0.29\hsize]{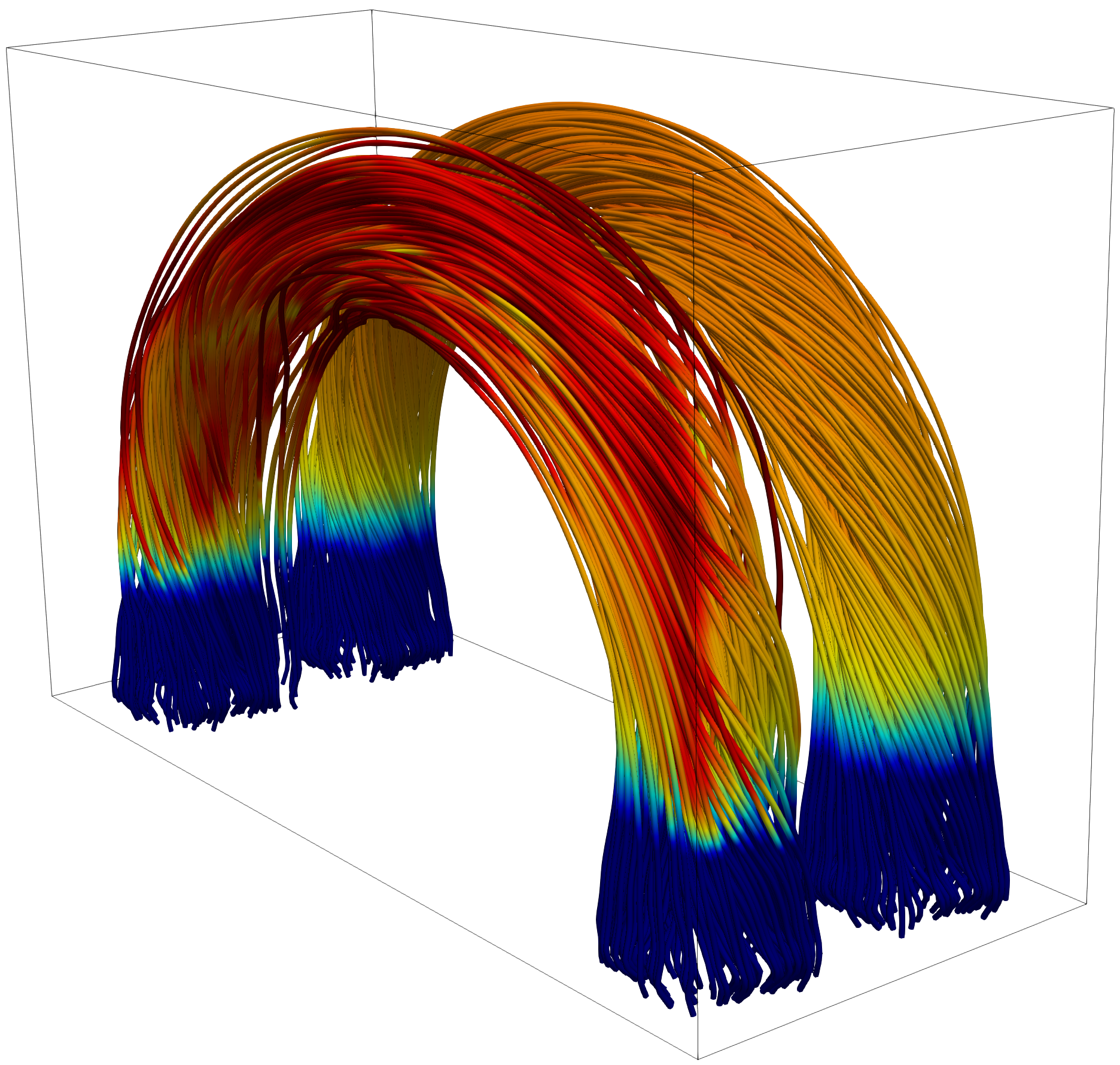}
  \includegraphics[width=0.29\hsize]{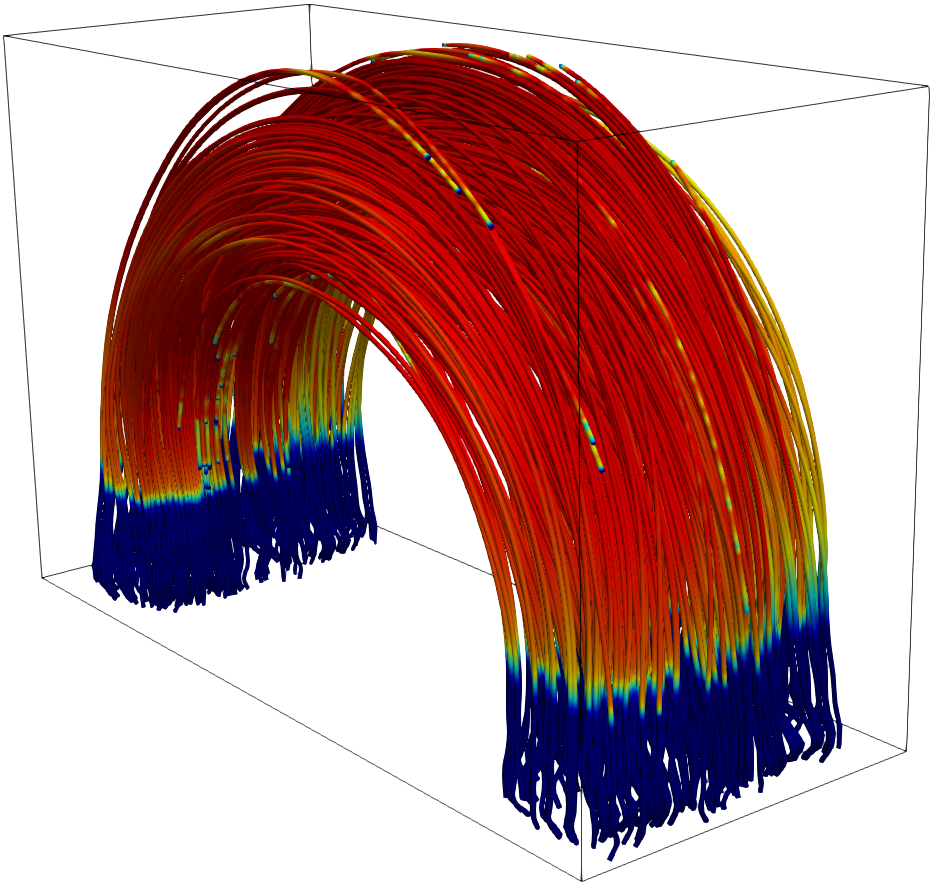}
  \includegraphics[width=0.08\hsize]{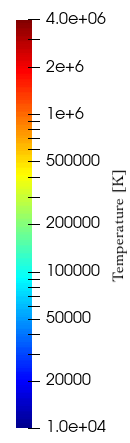}
  \caption{3D rendering of the magnetic field lines inside the box at the same times shown in Fig.\ \ref{Fig:avalance}. Magnetic field lines are color-coded according to the plasma temperature.}
\label{Fig:3d_loop}
\end{figure*}

The emission intensity $I$ \citep{boerner2012initial} from the modelled (optically thin) plasma expected to be measured with MUSE is:
\begin{equation}
    I (x,y) = \int_0^{\infty} \Lambda_f (T) \, \mathrm{DEM} (T) \, dT
,\end{equation}
where $\Lambda_f (T)$ is the temperature response function per unit pixel for the line, $f$, and $\text{DEM} (T) = n_H n_e \frac{dz}{dT}$ is the differential emission measure (with $n_e$ the free electron density). Then, $\Lambda_f (T)$ combines the emission properties of the plasma with the response of the instrument:
\begin{equation}
    \Lambda_f (T) = \int_0^{\infty} G(\lambda, T) \, R_f (\lambda) \, A_{\mathrm{pix}} \, d \lambda, 
\end{equation}
where $G(\lambda, T)$ is the plasma contribution function, $R_f (\lambda)$ is the spectral response of the f-th instrument channel, and $A_{\mathrm{pix}}$ is the area of a single MUSE pixel ($0.167"\times0.4"$).
The temperature response functions for the three MUSE channels are shown in Fig.\ \ref{Fig:MandA}. 
They are calculated using CHIANTI 10 \citep{del2021chianti} with the CHIANTI ionization equilibrium, coronal element abundances \citep{feldman1992elemental}, assuming a constant electron density of $10^9\,\mathrm{cm}^{-3}$, and no absorption considered. Each curve is multiplied by a factor to convert photons into data-numbers (DN).
Each curve peaks approximately at the temperatures listed in table \ref{table:1}.
The intensity of a single cell, labeled in the numerical grid by the indices $\mathrm{i,j,k}$, in units of $\mathrm{DN}\,\mathrm{s}^{-1}\,\mathrm{pix}^{-1}$, is equal to \citep{de2022probing}:
\begin{equation}
    F_{i,j,k} = n_{e \, i,j,k}^2(T) \, \Lambda_f(T) \, \Delta z, 
\end{equation}
with $n_e \simeq \frac{\rho}{\mu \, m_H}$, where $\rho$ is the mass density, $m_H$ is the hydrogen mass, and $\Delta z$ is the cell extent along a specific line of sight (LoS), such as $\hat{z}$. We assume a Gaussian profile for each line, at a fixed temperature and density. Therefore, each cell, with a local velocity of $v_{i,j,k}$ along the LoS, contributes to the overall line spectrum with the following single-cell line profile:
\begin{equation}
    f_{i,j,k} (v) = \frac{F_{i,j,k}}{\sqrt{2 \pi \sigma_T^2}} \exp{ \left[ -\left(\frac{v - v_{i,j,k}}{\sigma_T} \right)^2 \right] } 
    \label{Eq:Gaussian_prof}
,\end{equation}
where 
\begin{equation}
\sigma_T = \sqrt{\frac{2\,k_b\,T}{m_{\mathrm{Fe}}}}
\label{Eq:Sigma}
\end{equation}
is the thermal broadening and $m_{\mathrm{Fe}}$ is the iron atomic mass, since we will look only at iron lines.
The synthetic instrument response, namely, the integrated emission along the LoS is computed as:

\begin{equation}
I_{0}^{i,j} = I_{i,j} = \sum_k F_{i,j,k}.
\label{eq:zero_momentum}
\end{equation}

MUSE will measure the line profiles, from which we can derive the line Doppler shifts and the line widths. We can compute the first moment of the velocity distribution along the LoS:  
\begin{equation}
I_1^{i,j} = v_{i,j} = \frac{\sum_{k} F_{i,j,k} v_{i,j,k}}{I_0}
\label{eq:first_momentum}
,\end{equation}
which can be compared to the Doppler shifts measured along the LoS. We can derive the non-thermal contribution to the line width (measured after subtracting the thermal line broadening, $\sigma_T$, and instrumental broadening) from the second moment of the velocity:
\begin{equation}
I_2^{i,j} = \Delta v_{i,j} = \sqrt{\frac{\sum_{k} F_{i,j,k} (v_{i,j,k} - I_{1i,j})^2}{I_0}}.
\label{eq:second_momentum}
\end{equation}

\begin{figure*}[h!]
   \centering
  \includegraphics[width=\hsize]{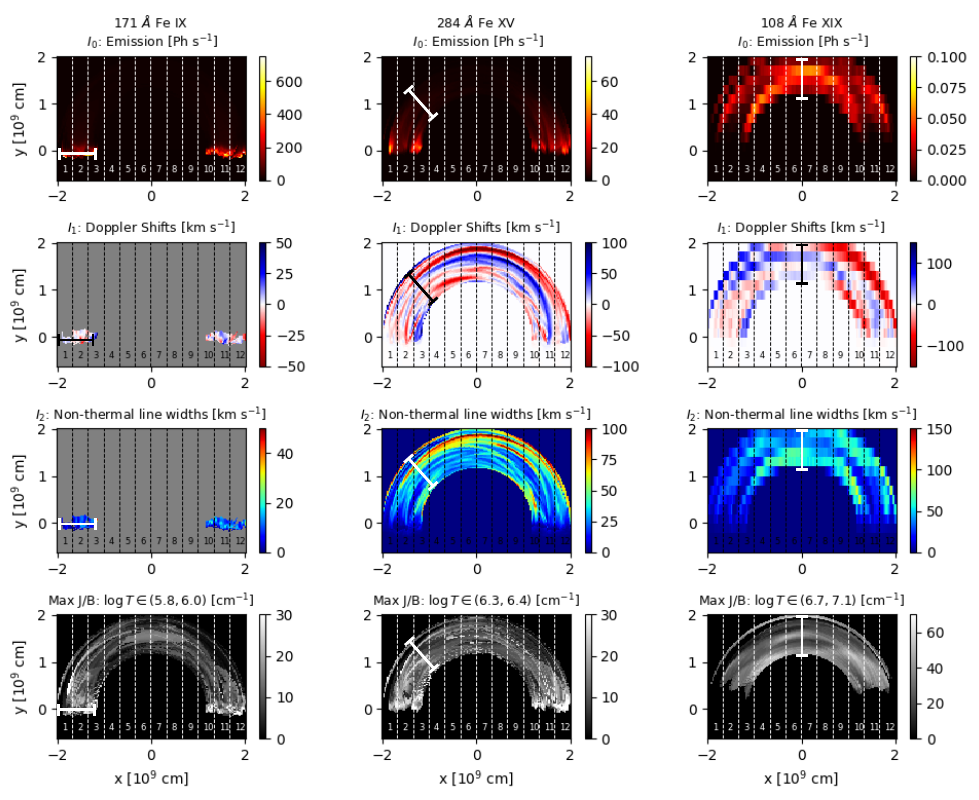}
  \caption{MUSE synthetic maps, from PLUTO 3D MHD model of the  multi-threaded magnetic flux tube discussed in Section \ref{sec:model}. Here we show the tube structure at time $\sim 180\,\mathrm{s}$ after the onset of the instability and with the loop in off-limb configuration (LoS along the $\hat y$ direction). From the top, the first row shows the intensity of \feix, \fexv, and \fexix\ emission lines.  Second row shows the related line shifts instead.  Third row  shows the non-thermal line broadenings. In the the \feix\ Doppler shifts and widths, we only show the  pixels where the line intensity exceeds the 5\% of its peak ($I_0 > 0.05 \, I_0^{\mathrm{max}}$). We show \fexix\ observables rebinned on macropixels ($0.4" \times 2.7"$). 
   Fourth row shows the maximum of the current density-magnetic field ratio (per each pixel plasma column) in three temperature bins, around the temperature peak of each line (see attached Movie 1). }
  \label{Fig:slits_los1_observables}
\end{figure*}

\section{Results}
\label{sec:results}

To replicate the typical, semicircular, coronal loops shape better, the original data have been remapped and interpolated onto a new cartesian grid. The method is discussed in Appendix \ref{sec:appendix}. Fig.\ \ref{Fig:3d_loop} shows the 3D rendering of magnetic field lines in the new geometry, at the same snapshot times used for Fig.\ \ref{Fig:avalance}. Magnetic field lines are color-coded according to the plasma temperature, ranging from $10^4\,\mathrm{K}$ in the chromosphere (blue-layer in the lower part of the box) up to $4 \times 10^6\,\mathrm{K}$ in the upper corona. The sequence clearly shows that both flux tubes are progressively heated from about 1 MK to more than 2 MK in the coronal part.

\subsection{Side view}
\label{sec:sfms}

\begin{figure}[h!]
   \centering
  \includegraphics[width=\hsize]{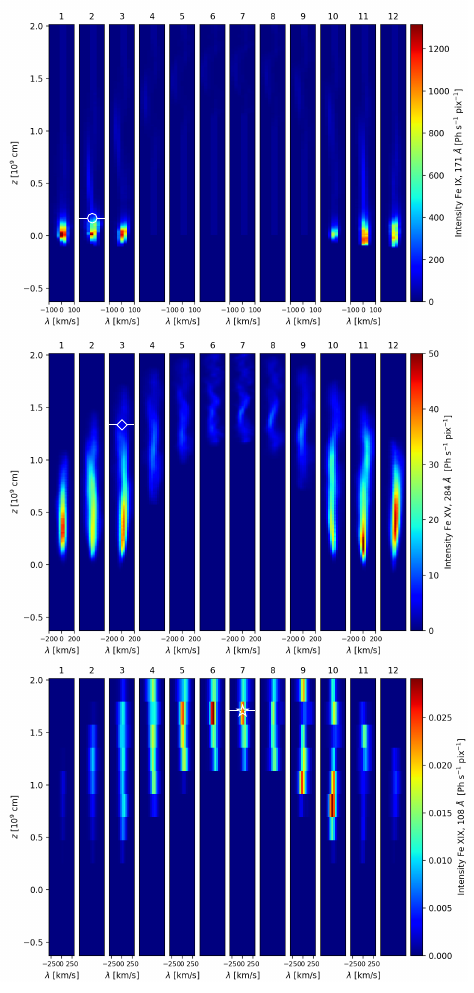}
\caption{MUSE synthetic line profiles for \feix, \fexv\ and \fexix\ emission line, as a function of height, $z$, slit number, and Doppler shift velocity, as seen in Fig.\ \ref{Fig:slits_los1_observables}. The spectral bins here are $\Delta v = 25\,\mathrm{km s}^{-1}$ for \feix, $\Delta v = 30\,\mathrm{km s}^{-1}$ for \fexv, and $\Delta v = 40\,\mathrm{km s}^{-1}$ for \fexix. The angular pixel along $\mathrm{z}$ is $0.4"$, equivalent to $290\,\mathrm{km}$ (see appendix \ref{sec:appendix}) for \feix\ and \fexv\ lines and $2.7"$ for \fexix. The positions of the profiles shown in Fig.\ \ref{Fig:line_profiles} are marked (circle, diamond and star).}
\label{Fig:Slit_intensity_lines}
\end{figure}

\begin{figure*}[h!]
   \centering
  \includegraphics[width=\hsize]{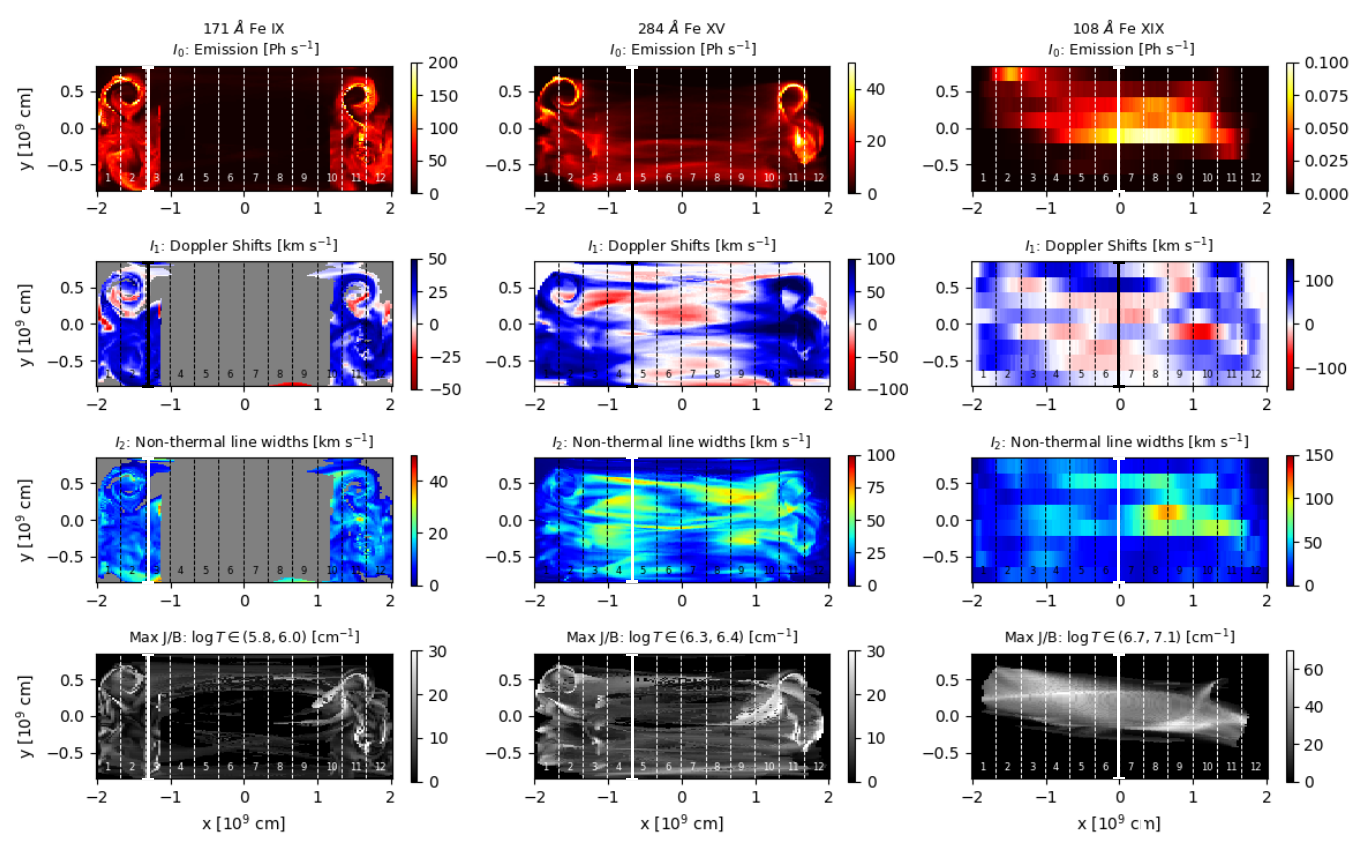}
  \caption{MUSE synthetic maps at time $\sim 285\, \mathrm{s}$ and with the LoS along the $\hat z$ direction (the loop is observed from the top). From the top, the first row shows the intensity of \feix, \fexv, and \fexix\ emission lines. Second row shows the Doppler line shifts. Third row shows the non-thermal line broadenings. In the the \feix\ Doppler shifts and widths we show only pixels where the line intensity exceeds the 5\% of its peak ($I_0 > 0.05 \, I_0^{\mathrm{max}}$). We show \fexix\ observables rebinned on macropixels ($0.4" \times 2.7"$). We also show the \fexix\ observables rebinned on macropixels ($0.4" \times 2.7"$). The fourth shows the maximum of the current density (per each pixel plasma column) in three temperature bins, around the temperature peak of each line. The 12 dashed vertical lines mark the position of the MUSE slits. 12 dashed vertical lines mark the position of the MUSE slits
  (see Movie 2). }
  \label{Fig:XY}
\end{figure*}

\begin{figure*}[h!]
   \centering
  \includegraphics[width=0.99\hsize]{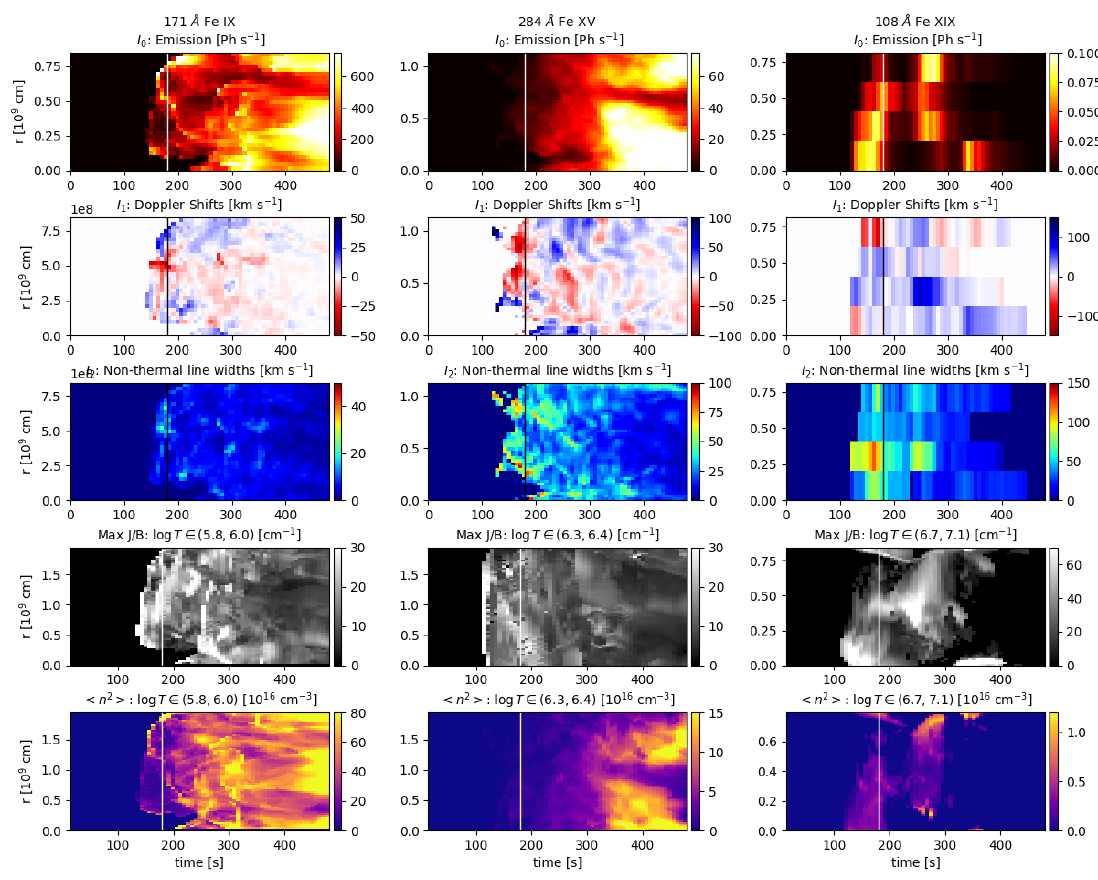}
  \caption{Same as in Fig.\ \ref{Fig:slits_los1_observables}, but considering the three single slits marked in Fig.\ \ref{Fig:slits_los1_observables}, as a function of time and of the distance from the loop's inner radius $r$.
  Times from Fig.\ \ref{Fig:slits_los1_observables} are shown as solid vertical lines.}
\label{Fig:multiple_filters_single_slit_rasting}
\end{figure*}

\begin{figure*}[h!]
   \centering
  \includegraphics[width=0.99\hsize]{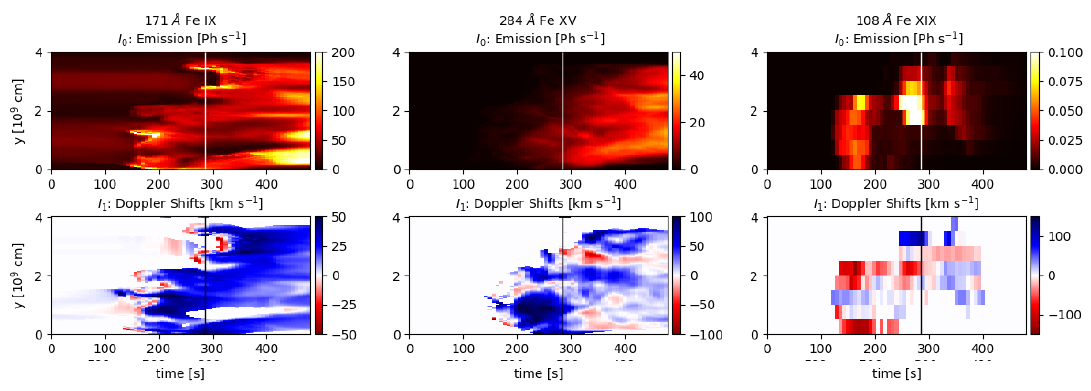}
  \caption{Same details as in the top two rows of  Fig.\ \ref{Fig:XY}, but considering three single slits (slit numbers 2, 4, and 6 in Fig.\ \ref{Fig:XY}, for the \feix, \fexv, and \fexix \ lines, respectively), as a function of time and $y$. The time  of Fig.\ \ref{Fig:XY} is marked (solid vertical lines).}
\label{Fig:XY_time_2rows}
\end{figure*}

Figure \ref{Fig:slits_los1_observables} shows a side view of the loop in the MUSE passbands, obtained from the PLUTO 3D MHD model of the kink-unstable, multi-threaded coronal loop (discussed in Section \ref{sec:model}) at the time $t = 180 \, \mathrm{s}$ (as in Fig.\ \ref{Fig:cut_1}).
The top three rows show the possible positions of the MUSE slits  (dashed vertical lines).
Each column for the MUSE line (\feix, \fexv,\ and \fexix) shows the intensity, Doppler shift, non-thermal line broadening, current density, and the emission measure integrated along the LoS and in a temperature range the line is most sensitive to. Here, we show MUSE observables only where the line intensity exceeds $5\,\%$ of the peak, as we address the main emission features. Regarding the \fexix\ line, we have decided to equally show emission features in this line for completeness, even though this line hardly reaches an acceptable level for detection with MUSE for this specific simulation, both because the response function is quite lower than the other two and because the densities are also lower at these temperatures. We expect more emission in this line for more energetic simulations, for instance, with a higher background magnetic field.

As shown in Fig. \ref{Fig:MandA}, the emission in the \feix\ line peaks around $1\,\mathrm{MK}$. Figure \ref{Fig:3d_loop} shows that the bulk of the flux tubes is soon heated to higher temperatures.  As a consequence, the emission in this line mostly comes from the transition region at the footpoints, where the temperature steadily remains around 1 MK.
On the other hand, the \fexv\ line is emitted higher in the flux tubes, although the loop is still not entirely bright at this time.
 
For the  hot \fexix\ line, we show the emission maps rebinned on macropixels ($0.4" \times 2.7"$),  increasing the level of emission closer to the detection level \citep{de2020multi}. We acknowledge that the emission in this line would come mostly from the region around the loop apex, where the temperature can reach around $10\,\mathrm{MK}$. 

The second row of Fig.\ \ref{Fig:slits_los1_observables} shows the Doppler shifts maps obtained as the first moment of the velocity distribution along the LoS (see Eq.\ \ref{eq:first_momentum}).
For the \feix\ line, we show only the regions where the emission is within 5\% of the maximum intensity, because we focus our attention on the brightest, low part of the loop. The overlapping velocity patterns along the LoS similar to those shown in Fig.\ref{Fig:cut_1} (bottom left) designate an irregularly alternate Doppler shift pattern at the footpoints visible in the \feix\ line, generally below $25\,\mathrm{km}\,\mathrm{s}^{-1}$, complicated by the starting instability.

The \fexv\ line captures the emission from most of the loop and emphasizes the chaotic filamentary (cross-section on the order of $1000\,\mathrm{km}$) pattern of alternating blue-red shifts ($\leq 50\,\mathrm{km}\,\mathrm{s}^{-1}$), due to the initial MHD instability. Within the limitations above, we might infer a similar pattern (with higher speeds to about $100\,\mathrm{km}\,\mathrm{s}^{-1}$) also in the \fexix\ line, although the structuring does not overlap with the \fexv\ one, because of different contributions along the LoS.

The third row shows maps of the second moment of each line, which is a proxy  of the non-thermal line broadening. As a reference, we mention that the thermal line broadenings as defined in Eq.\ \ref{Eq:Sigma}, at the line peak temperatures, namely, $T = 0.9\,\mathrm{MK}$ for the \feix\ line, $T = 2.5\,\mathrm{MK}$ for the \fexv\ line, and $T = 10\,\mathrm{MK}$ for the \fexix\ line, is $16\, \mathrm{km}\,\mathrm{s}^{-1}$, $27\,\mathrm{km}\,\mathrm{s}^{-1}$, and $54\,\mathrm{km}\,\mathrm{s}^{-1}$, respectively. The non-thermal broadening describes co-existing velocity components  in different directions along the LoS. In very elongated structures, the thermal and non-thermal components are comparable; however, for the \fexv\ (and \fexix\ ) line in outer shells, the latter can exceed the former significantly, with a chaotic structure resembling that of the Doppler shifts.  In the \fexv\ line, there is a long strip around the loop apex where the non-thermal broadening exceeds $75\,\mathrm{km}\,\mathrm{s}^{-1}$.
The fourth row shows the maximum of the current density-magnetic field ratio (per each pixel plasma column) in three temperature bins, around the temperature peak of each line, to locate the regions where reconnection is most likely to occur.

Figure\ \ref{Fig:Slit_intensity_lines} shows the synthetic line profiles for the \feix, \fexv, and \fexix\ lines, as a function of height $z$, slit number, and Doppler shift velocity, in a slit-like format, similar to the expected direct output of the spectrometer \citep[e.g.,][Fig.4]{de2022probing}. These profiles account for both the thermal and non-thermal line broadening since they have been obtained from the integration of the single-cell Gaussian profiles (see Eq.\ \ref{Eq:Gaussian_prof}) along the LoS and the angular coverage of each slit. 
The \feix\ line, bright at the footpoints only, does not show significant overall Doppler shift. The \fexv\ line, brighter low in the loop, shows some distortion when moving to the upper and fainter regions. In the faintest top region we also see an oscillatory trend. 

Regarding the \fexix\ line, we might say (solely in generical sense)  that we could expect to see some Doppler shifts that are more on the blue side, but only if we had enough sensitivity in the top loop region.
In general, having multi-slit allows us to see all these features across the loop in these three lines in the same instance.

\subsection{Top view}
\label{sec:fpb}

Figure\ \ref{Fig:XY} shows the MUSE synthetic images from a different perspective, with the LoS along $\hat z$ (i.e.m the loop is observed from the top) at times of $t = 285\,\mathrm{s}$, when the emission intensity is high in all three lines.
Also in this case, the maximum current density (weighted for the strength of the magnetic field) in the three lines temperature bins is shown in the fourth row, as in Fig.\ \ref{Fig:slits_los1_observables}.

In Fig.\ \ref{Fig:XY}, as expected, the \feix\ line intensity map shows bright footpoints at the sides, with no emission in between. In particular, helical patterns map the footpoint rotations of the two flux tubes.
The upper tube is better defined because not involved in the  kink instability yet. The related Doppler map shows mostly blue shifts of up-moving plasma, driven by the heating. At the footpoints, the non-thermal broadening is noisy and mostly below $20\,\mathrm{km}\,\mathrm{s}^{-1}$. Overall, it maps  the current density pattern rather well. 

The \fexv\ line is bright along most of the loop, but more so in the lower, kink-unstable loop than in the upper
kink-stable loop. Significant blue-shifts at about $100\,\mathrm{km}\,\mathrm{s}^{-1}$ are present low in the loop legs, marking bulk upflows. An alternation of blue and red shifts are present high in the loop. Non-thermal broadening above $50\,\mathrm{km}\,\mathrm{s}^{-1}$ is located at intermediate heights it and maps  the structuring of the current density quite well.

Also, from this perspective, the hot \fexix\ line is at its brightest around the loop apex and, in particular, at the boundary between the two flux tubes (in the mid plane). The Doppler shift pattern is diluted by rebinning, except for a couple of spots of redshift and non-thermal broadening.

\subsection{Time evolution}
\label{sec:mfss}

In Fig.\ \ref{Fig:slits_los1_observables}, we mark the cross-sections  along the loop where each line is bright and we use them to analyse the time evolution of the lines emission.
In Fig.\ \ref{Fig:multiple_filters_single_slit_rasting}, we show the same quantities as in the top three rows of  Fig.\ \ref{Fig:slits_los1_observables}, but considering three single slits (marked in Fig.\ \ref{Fig:slits_los1_observables}) as a function of time. The time of Fig.\ \ref{Fig:slits_los1_observables} ($t = 180\,\mathrm{s}$) is marked.

The \feix\ line, sampled at one loop footpoint, starts to brighten quite early ($t \sim 100\,\mathrm{s}$) overall and, in particular, in correspondence to the outer shell of the loop. While it does not occur uniformly, the brightness increases progressively over time. Conversely, the Doppler shift and the non-thermal broadening show higher values earlier in the evolution, around 150-$200\,\mathrm{s}$. Both of them have a patchy pattern, with alternating red and blue Doppler shifts and high small-line broadening, respectively.

The \fexv\ line brightens significantly after $t = 200\,\mathrm{s}$, which is quite later than the \feix\ line, because of the time taken by dense plasma to come up from the chromosphere. The brightening is more uniform than that seen for the \feix\ line and it already gets to its peak, in part of the loop,  at about $t = 300\,\mathrm{s}$. Also in this case we have more significant dynamics at early times (between $t = 100\,\mathrm{s}$ and $t = 300\,\mathrm{s}$), with Doppler shifts declining well below $50\,\mathrm{km}\,\mathrm{s}^{-1}$ and broadenings below $20\,\mathrm{km}\,\mathrm{s}^{-1}$ by the time the intensity peak is reached. However, the predicted intensities are already high enough \citep{de2020multi} at the time of peak shift or broadening to detect those line properties. In the Doppler-shift maps, we see more defined alternating red and blue patterns both in terms of the time and along the slit.

While it is quite faint, the \fexix\ line, emitted at the loop top, brightens very early ($t \sim 100$ s), for a short time and with an irregular pattern,  marking the time interval when the heating from reconnection is most effective (because the density is still relatively low) and the temperature has a peak \citep[see Fig.9 in][]{cozzo2023coronal}. In the interval between $100\,\mathrm{s}$ and $300\,\mathrm{s}$ we also see a bright spot of blue shift ($100\,\mathrm{km}\,\mathrm{s}^{-1}$) and one of line broadening  ($100\,\mathrm{km}\,\mathrm{s}^{-1}$.)

Figure\ \ref{Fig:XY_time_2rows} is similar to Fig.\ \ref{Fig:multiple_filters_single_slit_rasting}, but the evolution is taken along a line in the XY plane, with a LoS that is parallel to the Z-direction (top view). Also, in this case, we see an increasing line luminosity in the \feix\ and \fexv\ lines and flashes in the \fexix\ line. The evolution of the Doppler shifts cut shows clearly the presence of strong and long-lasting upflows from the loop footpoints (\feix).

Movie 1 shows the system evolution in the side view of Fig.\ \ref{Fig:slits_los1_observables}, from the onset of the instability to 500 s, when the system is close to a steady state. 
Significant currents appear first in the low temperature bin and around the loop apex at the onset of the instability and then again after 1 min in the other temperature bins. After about $2\,\mathrm{min}$, when the  plasma suddenly reaches $10\,\mathrm{MK}$, we see some faint emission at the top in the \fexix\ line, which rapidly extends down along the loop. At about $t=150\,\mathrm{s}$ we start to see emission also in the other two lines, starting from the loop footpoints, and then propagating upwards in the \fexv\ line only. At $t=300\,\mathrm{s}$ the whole loop appears illuminated in this line, while the \fexix\ line rapidly disappears completely. In the \fexv\ line, we clearly see bright fronts moving up and down and also that the brightness is not uniform across the loop, but with a filamented structuring which changes in time. 

The Doppler shift images have quite well defined patterns around $t=170\,\mathrm{s}$ (blue on the right, red on the left in the \fexv\ line), alternating wide strips in the \fexix\ line. These patterns rapidly become finely structured and chaotic by $t \sim 200$ s in both lines. In the first 2 min, we also see large intense (red) spots of non-thermal broadening in the same two lines, which  become filamentary later on, until they decrease significantly below $50\,\mathrm{km}\,\mathrm{s}^{-1}$ after about $5\,\mathrm{min}$.
The movie confirms that there is some level of correspondence between the evolution and structuring of the non-thermal broadening in the \fexv\ line and those of the current density.

When looking more closely at the single lines, the
\fexix\ line shows a brightening strip at the onset of the first instability ($130 s < t < 160 s$) at the low boundary around the apex of the curved tube. There we can also see some blue-shift, more to the left, a spot of non-thermal broadening ($ > 100$ km/s) and intense current density. These features are produced by the folding of elongated structures which align along the LoS. We see an analogous effect at the onset of the second instability ($250 s < t < 290 s$), this time involving more the top boundary.
The evolution of the \fexv\ line emission is more independent of the onset of the instabilities. In this side view, the loop starts to brighten from the footpoints after $t=150$ s and is filled in about 100 s. Then, its intensity increases more or less uniformly and progressively. Irregular Doppler-shifts and non-thermal broadening patterns are more intense at some times but progressively fade out.  
The \feix\ line shows irregular patterns, which are difficult to discern as they are very flat at the footpoints. 

Movie 2 shows the system evolution (analogously to Movie 1) from the top view of Fig.\ \ref{Fig:XY}. Again we clearly see the first brightening in the \fexix\ line in a thin strip at the boundary between the two rotating tubes ($t = 125\,\mathrm{s}$). Then the first unstable loop brightens up in all lines at the same time. At $t = 153\,\mathrm{s}$, bright helices appear at the footpoints in both \fexv\ and \feix\ lines and a broad oblique band extending over the whole loop length in the \fexix\ line. The \fexv\ and \feix\ brightenings rapidly extent to the other flux tube, with two definite helices. The helices rapidly blur and fade away in the \fexv\ line, while the emission gradually bridges over the whole loop region. Initially, it is brighter in the first unstable tube.
In the \fexix\ line, there are rapid faint flashes of few tens of seconds involving different, but relatively large coronal regions:  the first unstable tube earlier and  the other one later.
Regarding the Doppler shifts, this view clearly shows the expected strong upflows as the flux tubes initiate the instability, from $t \sim 200$ s. We see strong blueshifts in the first unstable tube at the footpoints in the \feix\ line, propagating to the body of the loop in the \fexv\ line, and continuing with redshifts in the \fexix\ line flows, moving from one side to the other.

Also, from this view, we see the strongest non-thermal broadenings soon after the instabilities are triggered in both flux tubes, with a structuring similar to the one shown by the current density, especially in the \fexv\ and \fexix\ lines. In particular, the \fexix\ line confirms the large non-thermal broadening when the line is more intense.
Looking in more detail at each line, in the \fexix\ line emission boundary strips brighten again, due to the folding of emission sheets that overlap along the LoS. At the same time, there are significant downflows and line broadening, downstream, or close to the areas of intense current density.

The \feix\ emission shows the helical patterns clearly at the onset of the instability ($t>150\,\mathrm{s} , t > 260 \,\mathrm{s}$). Alternating blueshifts and redshifts are also present at the onset of the instability ($t = 167s$, with velocity $v>50 $km/s), then  blue-shifts become dominant ($t>200$ s). Transient non-thermal broadenings are present for $160s <t<330s$ (3 min, $\Delta v \sim 30$ km/s). 

In the \fexv\ line, we see the ignition of the lower part of the first unstable tube at $t=167$ s, the whole tube is bright at $t=264$ s. The other tube starts to brighten  at $t=264$ s, and is entirely bright at $t=313$ s. Strong blueshifts appear at the footpoints early after each tube becomes unstable and extends upwards. Redshifts are present in area of weak emission, as well as large non-thermal broadenings.

\section{Discussion and conclusions}
\label{sec:conclusions}

This work addresses how coronal loops, heated by an MHD avalanche process, would be observed with the forthcoming MUSE mission.
It is an extension of the modeling-oriented work described in \cite{cozzo2023coronal} and it delves into the synthesis of spectroscopic observations in \feix, \fexv, and \fexix\ lines with the forthcoming MUSE mission, which will probe plasma at $\sim 1 \,\mathrm{MK}$, $\sim 2 \,\mathrm{MK}$, and $\sim 10 \,\mathrm{MK}$, respectively, at high spatial and temporal resolution.

In this work, forward-modeling bridges theoretical investigation of loop heating through MHD instability \citep{hood2009coronal,cozzo2023coronal} with realistic observations in the EUV band. This approach represents a significant progress both because it is based on a non-ideal MHD model (see also \citealt{guarrasi2014mhd}, \citealt{cozzo2023asymmetric}) and because it realistically accounts for future instrumental improvements (the MUSE mission). In particular, we provide specific observational constraints that are useful for testing the model and guide future modeling efforts. Forward-modeling for comparison with real observations requires a very detailed and complete physical description and including the space-time dependence at different spectral lines makes this task even more difficult. Historically, time-dependence has been possible using the 1D hydrodynamic modeling applied, for instance, to the light curves and line spectra of impulsively heated loops, \citep[e.g.,][]{peres1987hydrodynamic, antonucci1993simulations, testa2014evidence}. Constraints on the heating parameters of non-flaring loops were derived from comparing space-time dependent loop modeling with the brightness distribution and evolution observed with TRACE \citep{reale2000brightening1, reale2000brightening2} and SDO/AIA \citep{price2015forward}. Collections of single loop models have been used to synthesize the emission and patterns of entire active regions \citep{warren2007static, bradshaw2016patterns}. Collections of 1D loop models have been used also to describe the emission of multi-stranded loops \citep{guarrasi2010coronal}. 

Regarding proper multi-D MHD loop modeling, 
\cite{reale20163d} showed the synthetic emission of a coronal loop heated by twisting in two EUV (SDO/AIA $171\,\AA$ and $335\,\AA$, \citealt{lemen2012atmospheric,pesnell2012solar}) channels and in one X-ray (Hinode/XRT Ti-poly, \citealt{lee2009evolution}) channel. 
Also in that case, only the transition region footpoints brighten in the $171\,\AA$ channel since the loop plasma is mostly at temperatures around 2–3 MK, at which the 335 \AA\ channel is more sensitive. The central region of the coronal loop, at higher temperature plasma, is also bright in the X-ray band.

A list of preliminary applications of MHD forward-modeling to the MUSE mission is reported in 
\cite{de2022probing} and  \cite{cheung2022probing}.  
They  show that Doppler shifts and line broadening are important tools to provide key diagnostics of the mechanisms behind coronal activity. Moreover, they point out the importance of coupling high-cadence, high-resolution observations with advanced numerical simulations.

The full 3D MHD model in \cite{cozzo2023coronal} describes a release of magnetic energy in a stratified atmosphere. It accounts for a multi-threaded coronal loop made up by two (or more) interacting magnetic strands, subjected to twisting at the photospheric boundaries. The global, turbulent decay of the magnetic structure is triggered by the disruption of a single coronal loop strand, made kink-unstable by twisting. Massive reconnection episodes are triggered and determine local impulsive heating and rapid rises in temperature. These are then tempered by efficient thermal conduction and optically-thin radiative losses. The overpressure drives the well-known chromosphere evaporation into the hot corona and consequent loop brightening in the EUV and X-ray bands. 

The model allows for both space and time resolved diagnostics. In particular, we synthesized MUSE observations in the \feix, \fexv, and \fexix\ emission lines, sensitive to a wide range of emitting plasma temperature, i.e.,  $\sim 1\,\mathrm{MK}$, $\sim 2\,\mathrm{MK,}$ and $\sim 10\,\mathrm{MK}$, respectively.

The modeled loop is a typical active region loop of length 50 Mm with an average temperature of about 2.5 MK. As such, much of it its brightness is steadily in the \fexv\ line, only at the footpoints in the \feix\ line, and (very faintly and transiently) at the apex of the \fexix\ line.

\begin{figure}[h!]
   \centering
  \includegraphics[width=\hsize]{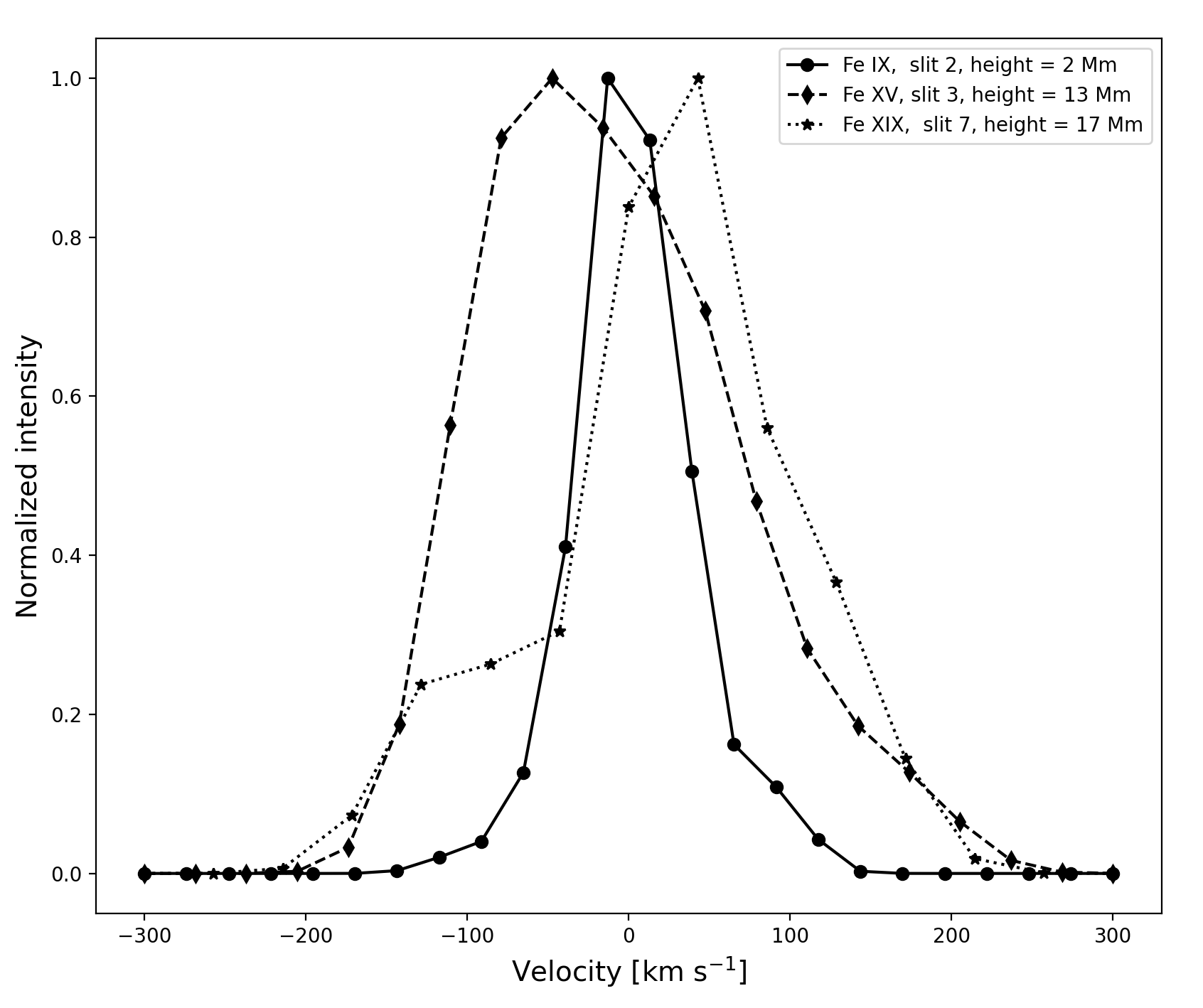}
  \caption{
  Detailed (normalized) profiles of the \feix \ (solid), \fexv \ (dashed), and \fexix (dotted) lines at the positions marked in Fig.\ \ref{Fig:Slit_intensity_lines}.
  }
\label{Fig:line_profiles}
\end{figure}

The loop is initially cold ($T \lesssim 1 \, \mathrm{MK}$) and tenuous ($n \sim 10^8 \, \mathrm{cm}^{-3}$), with faint emission only in the $171\,\AA$ channel of \feix.
Rapid, high temperature peaks, with faint emission in the \fexix\ line, occur during the dynamic phase of the instability, but most of the radiation is emitted by \fexv\ at a temperature of $\gtrsim 2 \,\mathrm{MK}$ (Fig.\ \ref{Fig:3d_loop}).

In Fig.\ \ref{Fig:Slit_intensity_lines}, we show the overall distribution of lines intensity and shape at time $\sim 180\,\mathrm{s}$.
In this case, line profiles show deviations from the Gaussian profile, such as an asymmetric shape and/or multiple peaks, as a consequence of the dynamic and impulsive behaviour of the instability. 
This is remarked also in Fig.\ \ref{Fig:line_profiles} with three examples   of \feix\ (solid), \fexv\ (dashed), and \fexix\ (dotted) profiles at different slits and heights. The  main peak of the 
\feix\ line  is almost at rest, whereas \fexv\ and \fexix\ show strong Doppler shifts (blue and red, respectively, at $|v|  \gtrsim 50 \mathrm{km s}^{-1}$).
In all these cases, there is an increase of non-thermal width in locations where reconnection seems to occur.
In particular, multi Gaussian components appear in \feix\ line (small, red-shifted peak at $v \sim 100 \mathrm{km s}^{-1}$), in the \fexix\ line (small, blue-shifted peak at $v \sim -100 \mathrm{km s}^{-1}$) and the \fexv\ line (strong contribution at rest).

\cite{testa2020coronal} show evidence of plasma heated at $10\,\mathrm{MK}$ during a microflare. In particular, they analyzed the coronal  ($131$ and $94\,\AA$ channels of AIA) and spectral (\fexxiii\ $263.76\,\AA$ line of Hinode/EIS) imaging and compared observations with hydrodynamic 1D modeling of a single loop heated by a 3 min pulse up to $12\,\mathrm{MK}$.
Forward modeling from the simulations provides additional evidence of the coronal loop multistructuring into independently heated substrands.

We have obtained results in both a spatially resolved (Section \ref{sec:sfms}) and a time-resolved (Section \ref{sec:mfss}) fashion. In all cases, 
we conclusively show how the three lines provide plasma information at different places, dynamical stages, and physical conditions. In particular, we have managed to efficiently disentangle the instability evolution into: (\feix) foot point response and plasma ablation in the transition region; (\fexv) over-dense and warm plasma rising at intermediate heights; and, faintly in this case, (\fexix) hot flaring plasma inside current sheets.

Progressive brightening is expected in the cooler lines, while we expect that the hottest line might be bright only occasionally and with a lower count rate. 
This is due to the low densities at the loop apex and the small filling factor, namely, only a small plasma volume gets close to the \fexix\ line formation temperature. We might expect more emission for higher temperature loops, as produced, for instance, by a stronger background magnetic field.
The hot \fexix\ shows faint emission here as soon as the flux tube becomes unstable, therefore, it may potentially become a signature of the initial phase of the instability.
\fexix\ emission is expected to be short-lived and around the loop top, 
to be compared with recent evidence of traces of very hot plasma (e.g., \citealt{ishikawa2017detection, miceli2012x}). 

In this analysis, we have reproduced expected line profiles integrated along possible lines of sight (Fig. \ref{Fig:Slit_intensity_lines}). In a limb snapshot view (Fig. \ref{Fig:slits_los1_observables}), MUSE observations might show alternating red-blue patterns along the loop, especially in the \fexv\ line. The non-thermal broadening might be also quite irregular with filamented structure. The line profiles taken in a side view might show an irregular trend across the loop in the hot lines. The top view (Fig. \ref{Fig:XY}) provides complementary information, helical patterns in the footpoints region, and systematic blue-shifts in the loop legs during the rise phase, but also possible faint one-way flows and filamented non-thermal patterns in the corona. More intense plasma dynamics from Doppler-shifts and non-thermal broadenings are expected during loop ignition, namely, just after the onset on the instabilities and avalanche process. Blue-shifts might be more persistent in the 1 MK line. Cross-structuring is ultimately linked to the assumptions on dissipation, in particular the resistivity determines the cross-size. The photospheric drive plays a role too close to the strands footpoints while it does not seem to influence cross-structuring in the upper corona.  MUSE observations might provide crucial information. The instability propagates with a delay of about 2 min, the timing depends on some assumptions (thickness of the flux tubes/rotation radius, rotation rate, field intensity).

The \fexix\ line (if detectable) might be a fair proxy of strong and dynamic current buildups such as current sheets. They form and rapidly dissipate during the earliest, most violent phase of the instability. At the same time (i.e., at the time of the disruption of the threads), the \feix\ footpoint emission provides complementary information about the status of the magnetic structure (Fig. \ref{Fig:cut_1}), suggesting a potential diagnostics role in terms of extrapolating information about the linear phase of the instability (Fig.\ \ref{Fig:XY}), as it might outclass the observational restriction imposed by the small counts-rate obtained with \fexix\ line \citep{de2020multi}.

After the early stages of the instability, \fexv\ emission returns information about the evaporation process triggered by the avalanche. This is the only case in which almost the whole coronal loop structure becomes visible to the instrument.

Doppler shifts and non-thermal line broadening can provide additional information about the plasma dynamics, confirming the strongly turbulent behaviour of the avalanche process, which is otherwise difficult to grasp only from emission maps analysis. Doppler-shift maps also contribute complementary information on the strength of the chromospheric evaporation nearby footpoints (Fig.\ \ref{Fig:XY}). Indeed, after the instability, the lines from the simulation broaden and strongly blue-shift, due to the strong chromospheric evaporation up-flows and in agreement also with \cite{testa2020coronal} forward modeling of 1D microflaring coronal loop.

According to the estimated uncertainties in centroid and line width determination discussed in \cite{de2020multi}, minimum line intensities of $\sim 100/150$ photons for \feix/\fexv, lines and $\sim 20$ photons for \fexix, are required to obtain the desired accuracy.
In our simulations, \feix, and \fexv, Doppler shifts, and non-thermal line broadening can be accurately estimated in the brightest regions (above $5 \%$ of the intensity peak) and with exposure times of $\gtrsim 5\,\mathrm{s}$. Emission in the \fexix\ MUSE line would perhaps be detectable only with particularly deep exposures (> 30 s) with our setup, but with shorter exposures for stronger magnetic field.

This work addresses in detail the observability of a 2 MK coronal loop system that is triggered by an MHD instability and described by a full 3D MHD model. To be specific, this forward-modeling approach is applied to the EUV spectrometry to be obtained with the forthcoming MUSE mission. Although calibrated on the high capabilities of this ambitious mission, our analysis is generally valid for observations made with instruments working in similar bands, such as present-day SDO/AIA, Hinode/EIS, Solar Orbiter SPICE \citep{anderson2020solar}, and EUI \citep{rochus2020solar}, or the forthcoming Solar-C/EUVST as well. The capabilities in terms of spatial, temporal, and spectral resolution of current observations (e.g., SDO/AIA, Hinode/EIS, Solar Orbiter SPICE and EUI) do not allow us to reach the diagnostic level shown by our analysis and further support the implementation of the MUSE mission.  

\begin{acknowledgements}
      GC, PP, and FR acknowledge support from ASI/INAF agreement n. 2022-29-HH.0. 
      This work made use of the HPC system MEUSA, part of the Sistema Computazionale per l'Astrofisica Numerica (SCAN) of INAF-Osservatorio Astronomico di Palermo.
      JR and AWH acknowledge the financial support of Science and Technology Facilities Council through Consolidated Grant ST/W001195/1 to the University of St Andrews.
      PT was supported by contract 4105785828 (MUSE) to the Smithsonian Astrophysical Observatory, and by NASA grant 80NSSC20K1272x.
\end{acknowledgements}

\bibliographystyle{aa} 
\bibliography{bibliography}

\begin{appendix} 
\section{Data interpolation}
\label{sec:appendix}

\begin{figure}[t]
   \centering
  \includegraphics[width=\hsize]{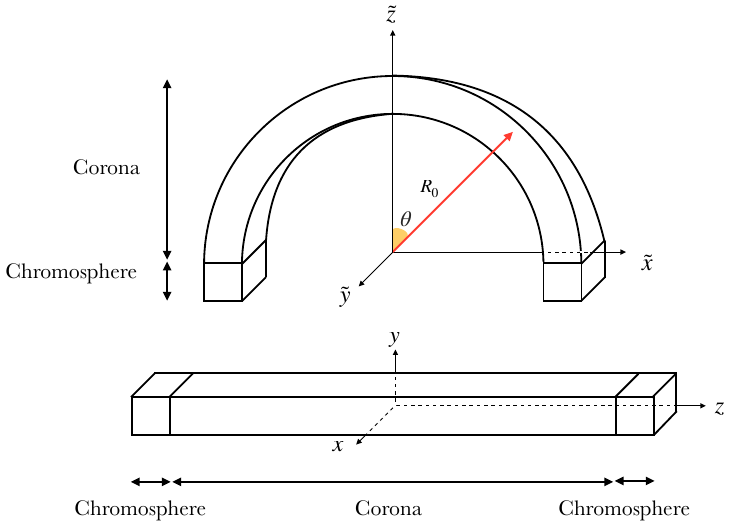}
  \caption{Original and interpolated box geometry. \textbf{Lower picture}: schematic representation of the 3D box containing the computational domain of the simulation. \textbf{Upper picture}: geometry of the interpolated domain mapping the straight flux tube into a curved one.}
  \label{Fig:Interpolation}
\end{figure}

To resemble the typical, semicircular, coronal loops shape, the original simulation results have been remapped and interpolated onto a new cartesian grid (see Fig.\  \ref{Fig:Interpolation}). In particular, points $(x,y,z)$ in the original grid corresponds to points $(\tilde x, \tilde y, \tilde z)$ defined by equations (\ref{eq:coordnates_1}) for the corona and (\ref{eq:coordnates_2}) for the chromospheric layer. 

Specifically, the coronal region of the domain ($z \in \left[-L, L\right]$) is transformed into a half cylinder shell with characteristic radius $R_0 = 2\,L/\pi$:
\begin{equation}
\begin{cases}
    \tilde x = R \sin \theta, \\
    \tilde y = x, \\
    \tilde z = R \cos \theta.
\end{cases} \hspace{1 cm} \mathrm{ for } \hspace{0.3 cm} z \in \left[-L, L\right]
\label{eq:coordnates_1}
\end{equation}
with $R = R_0 + y$ and $\theta = \frac{\pi}{2} \frac{z}{L}$.

The chromospheric layers are instead threaded as two parallel parallelepipeds of height $z_{\mathrm{max}} - L$:
\begin{equation}
\begin{cases}
    \tilde x = y, \\
    \tilde y = x,  \\
    \tilde z = L - |z|,
\end{cases} \hspace{1 cm} \mathrm{ for } \hspace{0.3 cm} |z| > L.
\label{eq:coordnates_2}
\end{equation}
New vectors components $(\tilde{v}_x, \tilde{v}_y, \tilde{v}_z )$, such as velocity and magnetic field, are obtained from the former $(v_x, v_y, v_z )$ using the following rule:
\begin{equation}
\begin{bmatrix}
    \tilde{v}_x \\
    \tilde{v}_y \\        
    \tilde{v}_z 
\end{bmatrix}
=
\begin{bmatrix}
0 & - \sin{\theta} & \cos{\theta} \\
1 & 0 & 0 \\
0 & - \cos{\theta} & \sin{\theta} 
\end{bmatrix}
\cdot
\begin{bmatrix}
    v_x \\
    v_y \\   
    v_z 
\end{bmatrix}
\end{equation}
where both triads $\tilde{v}_i$ and $v_i$ are functions of $(\tilde{v}_x, \tilde{v}_y, \tilde{v}_z )$.
In general, the correspondence between the former set of coordinates and latter is that the gravity vector $\mathbf{g}$ points uniformly along $\tilde z$.
Magnetohydrodynamic quantities are interpolated and rebinned in the new geometry at the given instrument resolution. In particular the grid pixel size, i.e. the dimension of the cells edges perpendicular to the LoS, will be equal to $170\,\mathrm{km} \times 290\,\mathrm{km}$, corresponding to $0.167" \times 0.4"$ i.e. the angular extension of the instrument's pixel (parallel and perpendicular to the slits orientation).

Figure\ \ref{Fig:3d_loop}  shows the resulting field line configuration colored by temperature in the new geometry at the three times already shown in Fig.\ \ref{Fig:avalance}.

\end{appendix}

\end{document}